\documentclass[usenatbib]{basi}
\usepackage[british]{babel}
\usepackage[varg]{txfonts}
\usepackage{rotating}
\usepackage{dcolumn}

\begin{document}
\def\lsim{\lower.5ex\hbox{$\; \buildrel < \over \sim \;$}}
\def\gsim{\lower.5ex\hbox{$\; \buildrel > \over \sim \;$}}
\def \simeq{\lower.3ex\hbox{$\; \buildrel \sim \over - \;$}}
\def\ch{\lower-0.55ex\hbox{--}\kern-0.55em{\lower0.15ex\hbox{$h$}}}
\def\lh{\lower-0.55ex\hbox{--}\kern-0.55em{\lower0.15ex\hbox{$\lambda$}}}
\def\eg{{\it e.g.,} }
\def\etal{{\em et al.} }
\def\ie{{\em i.e.,} }
\def\ee{{$e^--e^+$}}
\def\ep{{$e^--p^+$}}
\title[Relativistic fluids]{Effect of equation of state and composition on relativistic flows.}
\author[I. Chattopadhyay et~al.]%
       {I. Chattopadhyay$^1$\thanks{email: \texttt{indra@aries.res.in}},
       S. Mandal$^{2}$, H. Ghosh$^{3}$, S. Garain$^{3}$,  R. Kumar$^{1}$, and D. Ryu$^4$\\
       $^1$ARIES, Manora Peak, Nainital 263129, Uttarakhand, India\\
       $^2$IIST, Tiruvenantapuram, India\\
       $^3$S. N. Bose National Centre for Basic Sciences, Saltlake JD Block, Kolkata 700098\\
       $^4$Dept of Astronomy \& Sp. Sc., Chungnam National University, Kung Dong, Daejeon, South Korea}

\pubyear{2011}
\volume{00}
\pagerange{\pageref{firstpage}--\pageref{lastpage}}

\date{Received \today}

\maketitle
\label{firstpage}

\begin{abstract}
The thermal state of the fluid is governed by the ratio of the thermal and the
rest energy. This brings the composition of the
fluid into the picture. Although, fluid composed of lighter particles
(e.g: electron-positron pair plasma) at same temperature, is more relativistic compared to fluids
with finite baryon loading, but this is not necessarily true when baryon poor transonic fluid are compared
with each other. It can be shown that the transonic
pair-fluid is the least relativistic. This has far reaching consequences on accreting flows
around compact objects and are expected to have similar effect 
on relativistic outflows and explosive events as well.
\end{abstract}


\section{Introduction}\label{s:intro}

Relativistic fluids are expected in accretion discs around compact objects,
relativistic jets around AGNs, microquasars and in Gamma Ray Bursts.
A fluid is coined relativistic if its bulk velocity is comparable to the
speed of light ($c$). Terminal speeds of jets around microquasars and AGNs are observed to be close to $c$
\citep{1994Nat...371...46M, 1995ApJ...443...35Z}. In case of GRBs ultra-relativistic jets with
Lorentz factors $\gamma >$few${\times 100}$ are invoked to explain observations \citep{2002ARAA...40...137M}. General relativity requires that matter plunges onto the black hole
with the speed of light. 
Moreover, analysis of high energy photons received from micro quasars and AGNs, gives evidence of very high proton temperatures $T_p \gsim
10^{12} K$ and electron temperatures $T_e \gsim 10^9 K$.
If the thermal energy of the fluid is comparable to its rest energy ($kT/mc^2 \gsim 1$)
then the fluid is considered to be {\it thermally relativistic} and the thermal state is represented by the 
adiabatic index $\Gamma \sim 4/3$. Conversely, for
$kT/mc^2 << 1$, the fluid is called thermally non-relativistic ($\Gamma = 5/3$). Here $k$ and $m$
are the Boltzman constant and the mass of the fluid particles, respectively.
It is clear that the adiabatic index of the fluid depends on the thermal energy and is not
a constant. Constancy of $\Gamma$ is due to the non-relativistic treatment of the motions of the gas particles. A proper relativistic treatment results in a relativistic equation of state (EoS) of the fluid, and
a temperature dependent $\Gamma$ with correct asymptotic limits \citep{1938C, 1957S}. 
The relativistic EoS were applied to relativistic flows \citep{bm76, f87, metal04, metal05, rcc06, cr09, cc11},
although the general trend had been to use the fixed $\Gamma$ EoS (hereafter abbreviated as ID for ideal EoS).

In this paper, we present the applications of relativistic thermodynamics onto astrophysical fluids, and show that
consideration of relativistic EoS produces significantly different result from that of the ID EoS. Moreover, we will further show that
consideration of the composition of the fluid influences the flow solutions both qualitatively and quantitatively.

\section{Numerical Simulation with single species fluid}

\begin{figure}[h]
\centerline{\includegraphics[width=6.5cm]{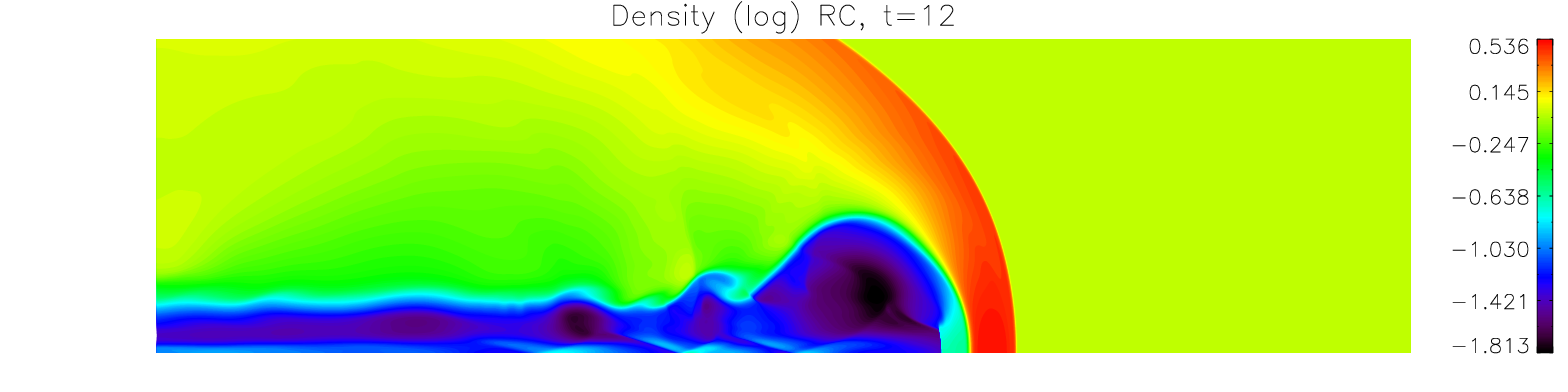} \qquad
            \includegraphics[width=6.5cm]{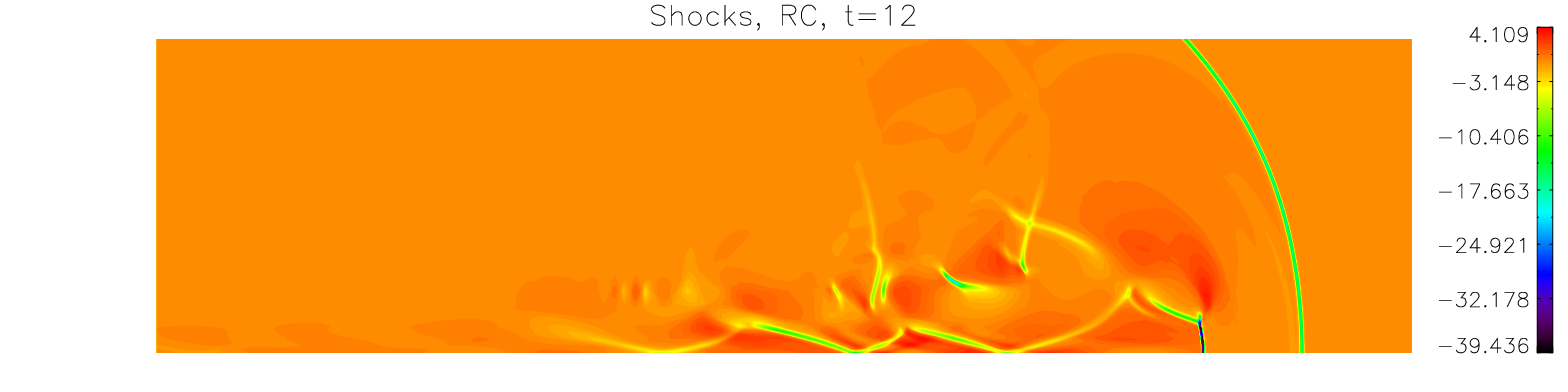}}
\medskip
\centerline{\includegraphics[width=6.5cm]{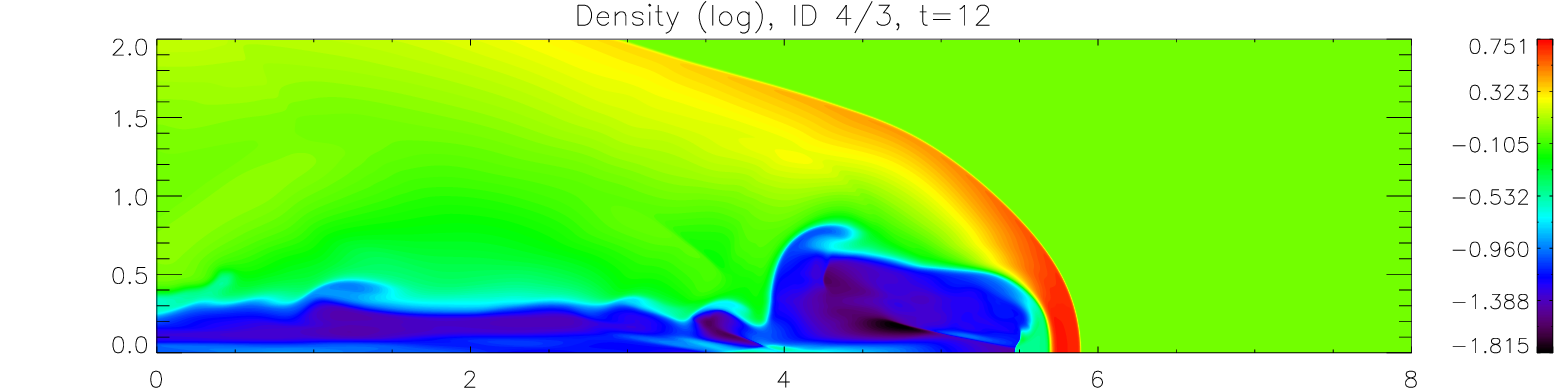} \qquad
            \includegraphics[width=6.5cm]{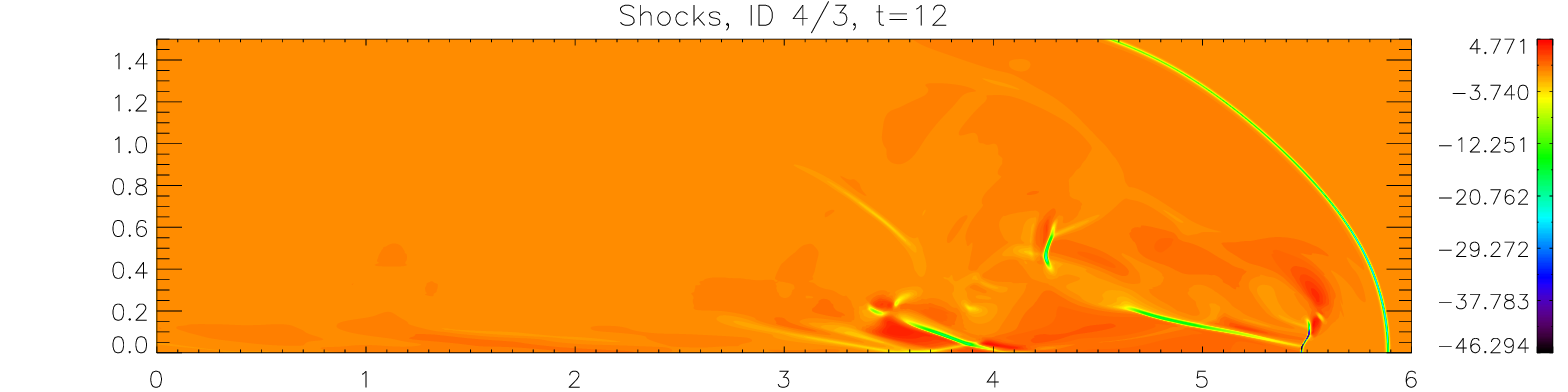}}
\caption{2D Slab jet simulation in Cartesian space-coordinates, resolution $1024\times256$, $r_b=12$ cells, $v_b=0.95$, $\rho_b/\rho_a=0.1$, and $p=0.01$. 
Density contours for RC (Upper left) and for ID EoS (Lower left) in log scale. The suffix `a' and `b' stands for ambient
medium quantities and jet beam quantities, respectively.
Projection of shock surfaces for RC EoS (upper right) and for ID EoS (Lower right).
All the figures are for the same snap shot t=12. \label{f:f1}}
\end{figure}

Assuming single species fluid and relativistic kinetic theory of gas, the EoS is computed by \citet{1938C,1957S}, which is
dependent on the ratios of modified Bessel functions of second and third kind.
A very good algebraic approximation (abbreviated as RC) was proposed by \citet{rcc06}, hereafter RCC06,
and is given by
\begin{eqnarray}
e=\rho h-p=\rho+p\left(\frac{9p+3\rho}{3p+2\rho}\right).\label{rceos}
\end{eqnarray}
Using RC EoS, a numerical simulation code was developed by RCC06 following 
the TVD formalism \citep{h83}. The relativistic equations of motion
are
\begin{equation}
 T^{\mu \nu}_{; \nu}=0 ~~~~ \mbox{and} ~~~~ (nu^{\nu})_{; \nu}=0,\label{eqmot1}
\end{equation}
The complete eigenstructure for Eqns \ref{rceos},\ref{eqmot1} were presented in RCC06. 
The first simulation is for 2D slab jet with RC and ID EoS
(see, caption of Fig. 1 for details).
Although the fluid with different EoS are launched with the same initial conditions, the evolution
of the jets are different, and more importantly, the shock structures generated by the two jets
due to the interaction with the ambient medium, are completely different. Such remarkable difference in shock structures
and overall flow variables,
will definitely generate different spectra.

\begin{figure}[h]
\centerline{\includegraphics[width=6.5cm]{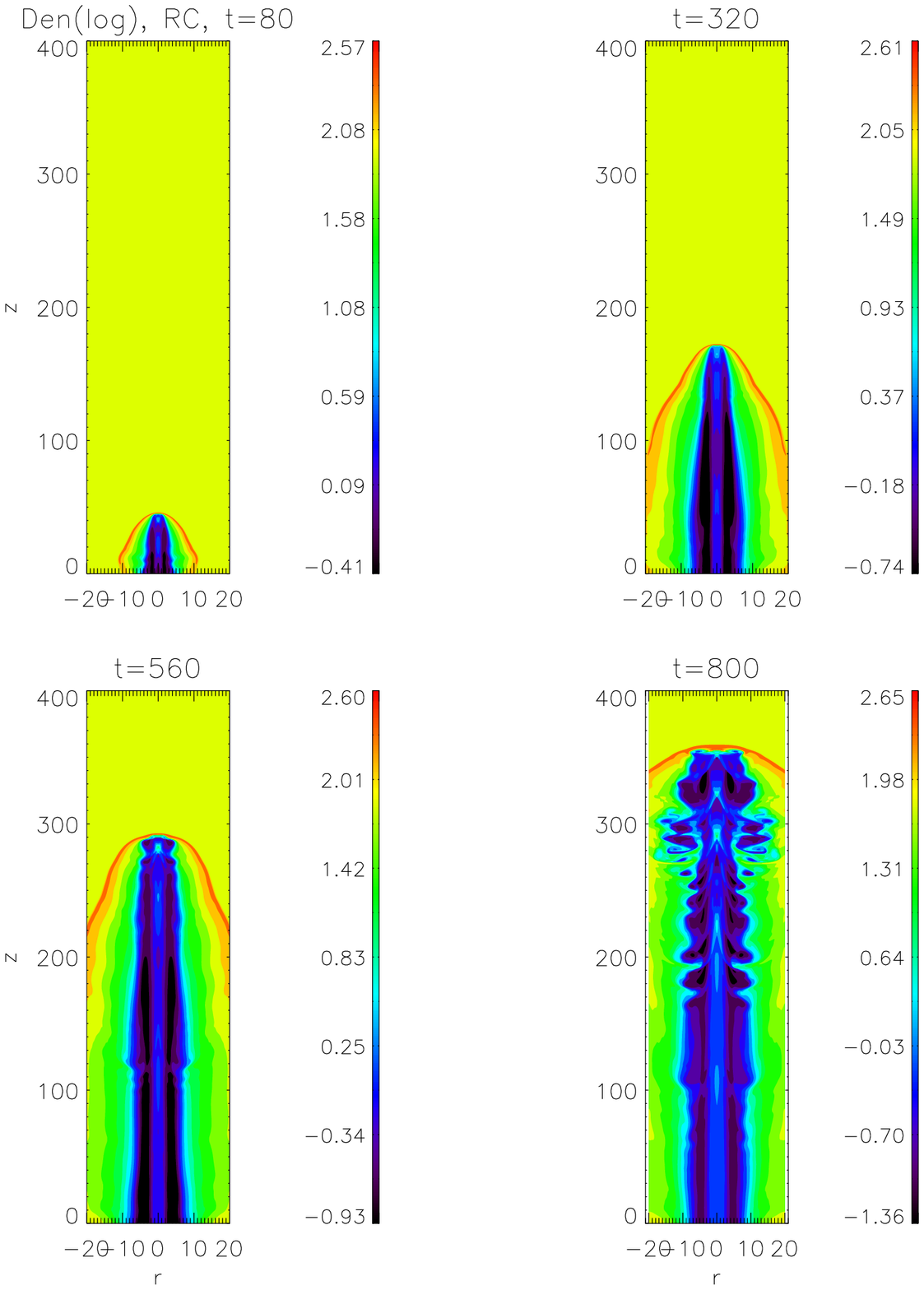} \qquad
            \includegraphics[width=6.5cm]{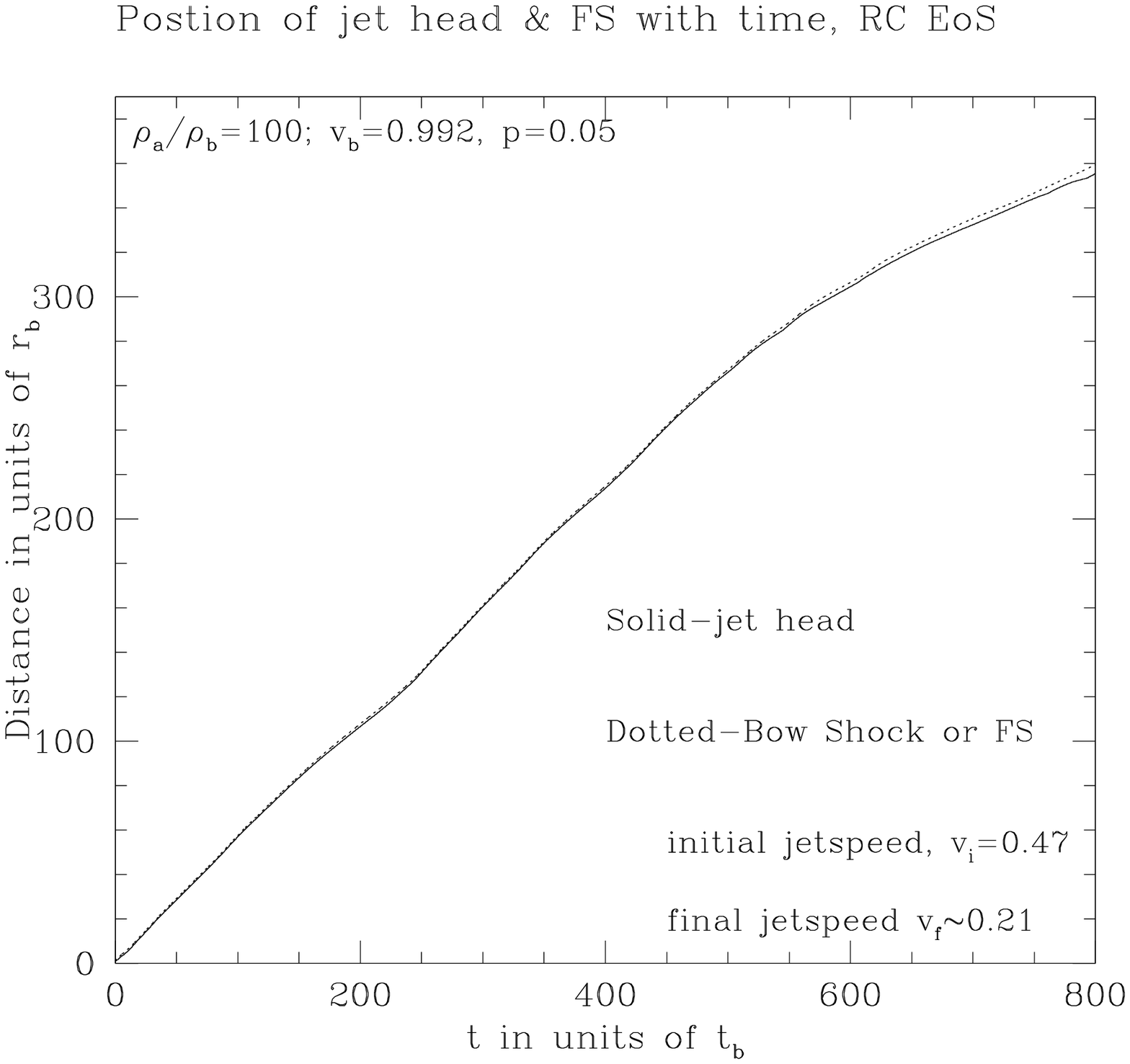}}

(a) \hskip 10.0cm (b)
\caption{2D cylindrical axisymetric jet simulation, resolution $240\times4800$, $r_b=24$ cells,
$v_b=0.992$, $\rho_b/\rho_a=10^{-2}$, and $p=0.05$. (a) Density contours (log scale) in the
$r-z$ plane for RC EoS, at four
time snaps. The unit of length is initial jet beam width $r_b$ and that of time is $t_b=r_b/c$ or the light crossing time of $r_b$.
(b) The time evolution of the forward shock (FS) and the jet head for the above case. \label{f:f1}}
\end{figure}

In Fig. 2a, four snap shots of the evolution of density contours for 2D cylindrical axisymetric jet 
simulation with RC EoS are plotted. In Fig. 2b, the forward shock
and jet head is plotted with time. As long as the jet accelerates, very little flow
structure is formed ($t\lsim500~t_b$), but as the jet
decelerates, \ie as the beam looses its kinetic energy, the back flowing material interacts with the beam itself
and many structures are seen to be formed.

\section{Multi species fluid}
\begin{figure}[h]
\centerline{\includegraphics[width=7.5cm]{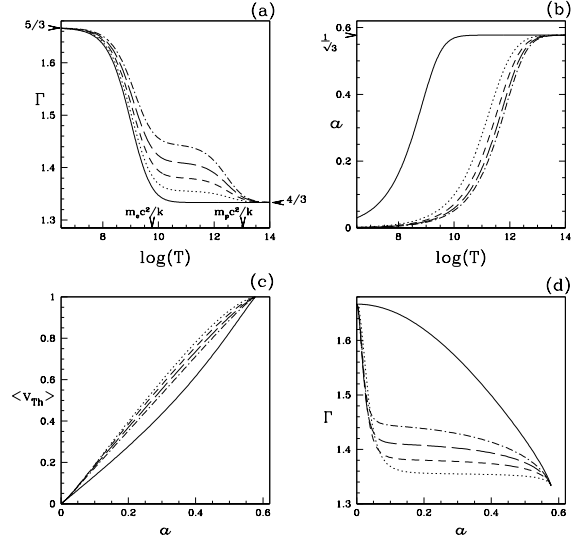}}
\caption{(a) $\Gamma$, and (b) sound speed $a$ are plotted with $log(T)$, (c) thermal velocity 
$<{\rm v}_{Th}>$ and (d) $\Gamma$ with
$a$, for $\xi=0$ \ie \ee (solid), $\xi=0.25$ (dotted), $0.5$ (dashed), $0.75$ (long dashed), and $1.0$ \ie \ep (dashed-dotted). }
\end{figure}

As has been noted in the previous section that a fluid is considered thermally relativistic if
$kT/mc^2\gsim 1$,
therefore, it is not only the $T$ alone but the ratio $T/m$, which
determines the thermal state of a fluid. Fluids composed of similar particles
are thermodynamically similar, and hence, their time evolution will be similar under similar physical
conditions.
However, a fluid composed of a mixture of dissimilar particles will be different.
Assuming a fluid composed of electron ($e^-$), positron ($e^+$) and protons ($p^+$),
where each species of the fluid obey RC EoS, then the proper energy density
is given by \citet{cr09}
\begin{equation}
 e=n_{e^-}m_ef, \label{eden}
\end{equation}
where,
\begin{equation}
 f=(2-\xi)\left[1+\Theta\left(\frac{9\Theta+3}{3\Theta+2}\right)\right]
+\xi\left[\frac{1}{\eta}+\Theta\left(\frac{9\Theta+3/\eta}{3\Theta+2/\eta}\right)\right]. \label{feq}
\end{equation}
In Eqn. \ref{feq}, $\Theta=kT/m_ec^2$ where, $k$, and $m_e$ are the Boltzman's constant
and electron rest mass, respectively. Here, $\xi$ is
the ratio of the number densities of protons and electrons, so $\xi=0$ is electron-positron
plasma or $e^--e^+$, and $\xi=1$ is electron-proton plasma or $e^--p^+$. Moreover, $\eta$
is the ratio between an electron and a proton mass. The definition of the adiabatic index $\Gamma=1+1/N$, where
$2N=df/d\Theta$. In Fig. 3a, $\Gamma$ is plotted as a function
of $T$. The thermal state of the fluid distinctly depends on composition parameter $\xi$
and $T$. At a given
$T$, the fluid composed of lightest particles (\ee) is more relativistic \ie $\Gamma$ is the lowest,
but $\Gamma$ increases with $\xi$. For non-relativistic
temperatures $T \lsim 10^7$K,
and ultra relativistic temperatures $T\gsim 10^{13}$K the dependence of the fluid thermodynamics on $\xi$ is nominal.

\begin{figure}[h]
\begin{center}
 \includegraphics[scale=0.35]{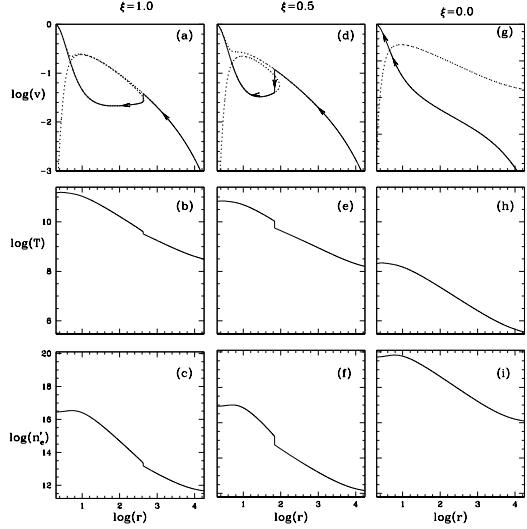}
\end{center}
\centerline{}
\caption{Constants of motion $\{{\cal E},\lambda\}=\{1.00009,3.4\}$.
The first column is for $\xi=1.0$ (Figs: a, b, c); the second for $\xi=0.5$ (Figs: d, e, f)
and the third for $\xi=0.0$ (Figs: g, h, i). Each row represents flow variables $log(v)$
(Figs: a, d, g);
$log(T)$ (Figs: b, e, h); and $log(n^{\prime}_e)$ (Figs: c, f, i); where $n^{\prime}_e=n_{e^+}+
n_{e^-}$.
Vertical jumps represent standing shocks. For $\xi=1.0$ $x_s=423.46$, and $\xi=0.5$ $x_s=68.78$. Solid curve
with arrow, the top row show solutions chosen by the shocked accreting matter from all the
associated solutions (dotted).}
\end{figure}

In Fig. 3b, the local sound speed $a=\Gamma p/(e+p)$ is plotted with $log(T)$. At same $T$, the fluid composed of lighter particles have higher sound speed. At the same thermal energy the fluid composed of lightest particles will be the most relativistic. The thermal
velocity $<{\rm v}_{Th}>$ is plotted with $a$ in Fig. 3c. The thermal velocity
$<{\rm v}_{Th}>$ can be easily expressed in terms of $\Theta$ and $\xi$ from the following relationship,
\begin{equation}
 e=\rho c^2 <\gamma_{Th}>={\sqrt{<q_{Th}>^2c^2+<m>^2c^4}}=n_{e^-}m_ec^2f, \label{eqvth}
\end{equation}
where, $<\gamma_{Th}>$, $<q_{Th}>$ and $<m>$ are the average particle Lorentz factor, particle momentum and particle rest mass in the fluid rest frame, respectively.  
Clearly, ${\rm v}_{Th}$ is minimum for the \ee fluid (solid), but is not maximum for \ep (dashed-dotted).
For a given value of $a$, ${\rm v}_{Th}$ is maximum for a fluid of $\xi\sim0.24$. The relation between $\Gamma$ and
$a$ is plotted in Fig. 3d. At a given $a$, $\Gamma$ is the least relativistic for \ee, and most relativistic for a fluid of $\xi\sim 0.24$.
Figure 3c shows that
at a given $a$, the momentum transferred by the lighter particles in the fluid frame
is far less than the heavier particles. As a result, $T$ of \ee fluid
is much less compared to that of the fluid with $0<\xi\leq1$. However, with the increase of $\xi$,
the rest energy per oppositely charged particles of the fluid increases too.
Hence, a fluid becomes most relativistic at the optimal value of $\xi \sim 0.24$.
\subsection{Single temperature, adiabatic flow onto black holes}
\begin{figure}[h]
\centerline{\includegraphics[width=4.5cm]{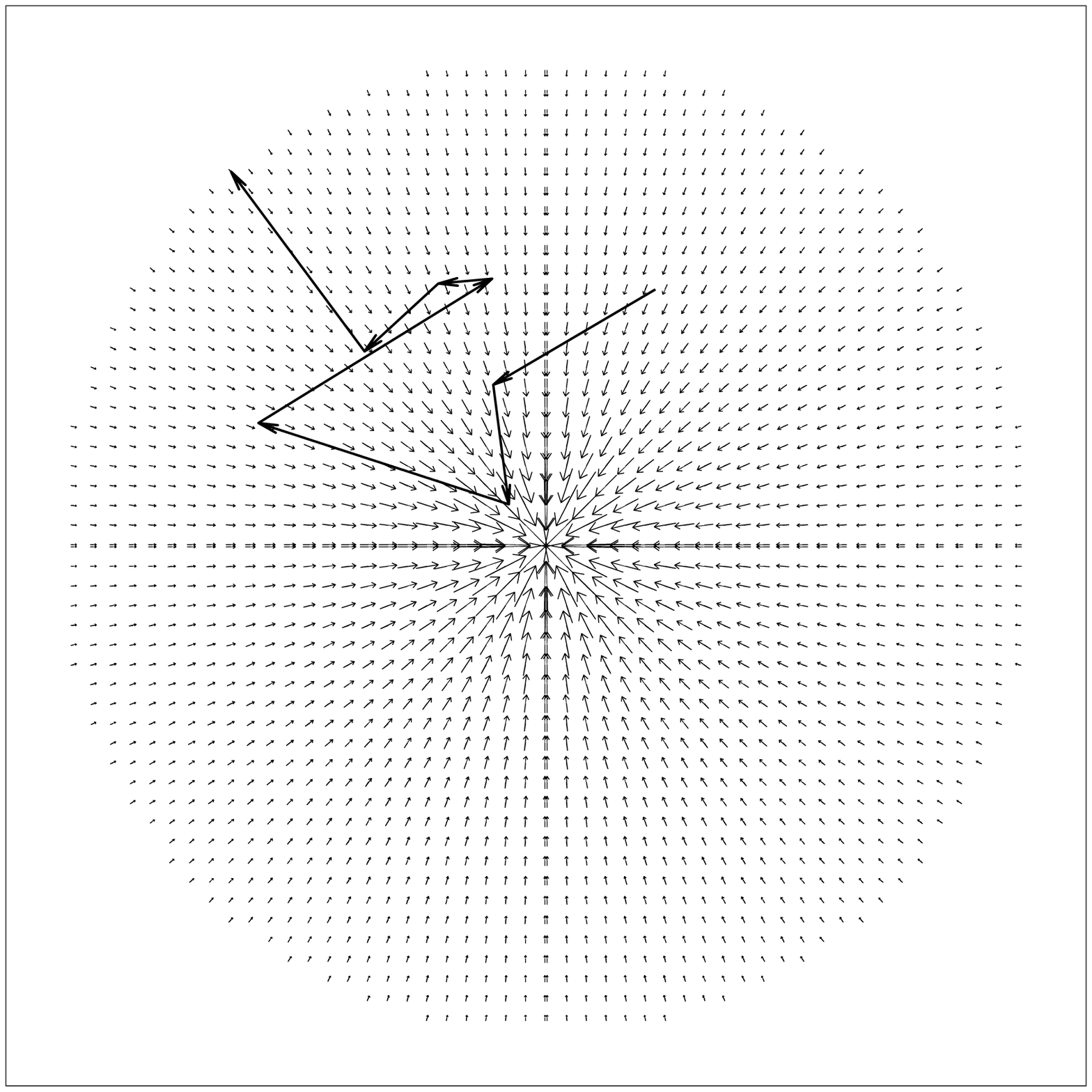} \qquad
            \includegraphics[width=4.5cm]{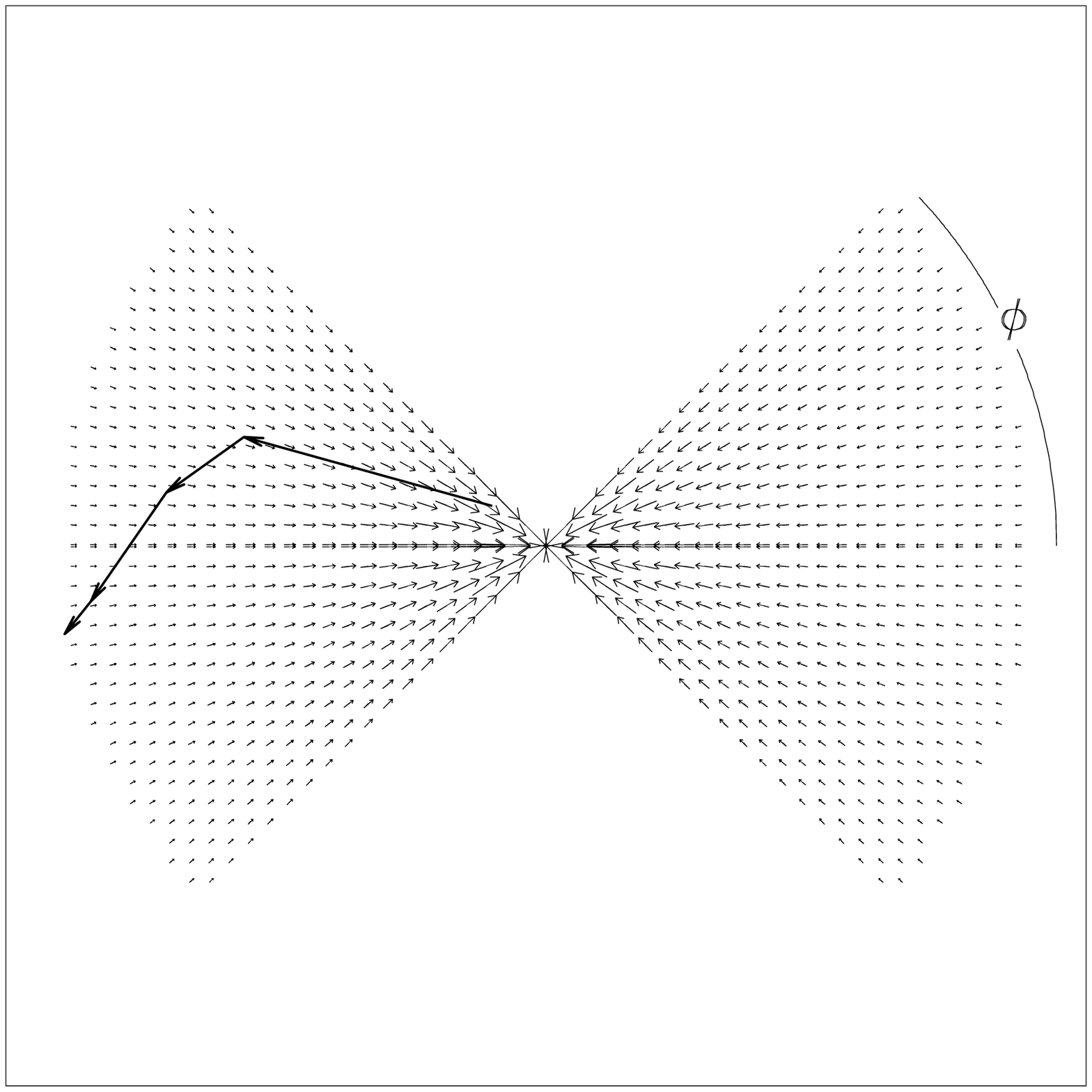}}

\hskip 2.0cm (a) \hskip 8.0cm (b)
\caption{2D projection of Monte-Carlo Simulation: (a) Velocity vector and path of a single photon for spherical accretion.
(b) Same for a conical flow. The composition is $\xi=1$. The simulation box is $500 r_g$. \label{f:f1}}
\end{figure}

Equations \ref{eqmot1}, when applied for steady, radial flow onto Schwarzschild black holes,
admits critical point relations \citep{c90, cr09}, which are also the sonic points, and are given by,
\begin{equation}
 v^2_c=a^2_c; \qquad a^2_c=\frac{1}{(2r_c-3)}, \label{eqradson}
\end{equation}
where, the subscript `$c$' denotes the quantities evaluated at the critical point $r=r_c$,
and $v^2=-u_ru^r/u_tu^t$ is the radial three velocity. A black hole accretion is definitely
transonic, so formation of a sonic point is a necessary feature of black hole accretion.
Two transonic solutions can be compared at the same sonic point,
or having the same specific energy at infinity, \ie same ${\cal E}|_{\infty}$.
Flows with same ${\cal E}|_{\infty}$ implies comparison of flows with same $a_{\infty}$. Moreover, flows with same $r_c$, implies
flows with same $a_c$. Hence according to Figs. 3b-3d, a transonic \ee fluid is expected to be the least
relativistic, and of the least temperature of all. \citet{cr09} showed it is indeed so. Since hot, rotating relativistic flows around black holes are supposed to form multiple critical points
and shock, then it should be interesting to see how composition affects the solutions of such flows. In Figs 4a-i, various flow variables are plotted, which show that for the same flow parameters
$\{{\cal E},\lambda\}=\{1.00009,3.4\}$, shock and multiple sonic points may form for flow with some baryon content, while for purely leptonic flow neither multiple sonic point form, nor shocks form \citep{cc11}.
Clearly, if the flow starting with same outer boundary condition can have such a widely different solutions,
then the radiative properties will be distinctly different.

\subsection{Monte Carlo simulation and spectrum}
\begin{figure}[h]
\centerline{\includegraphics[width=5cm]{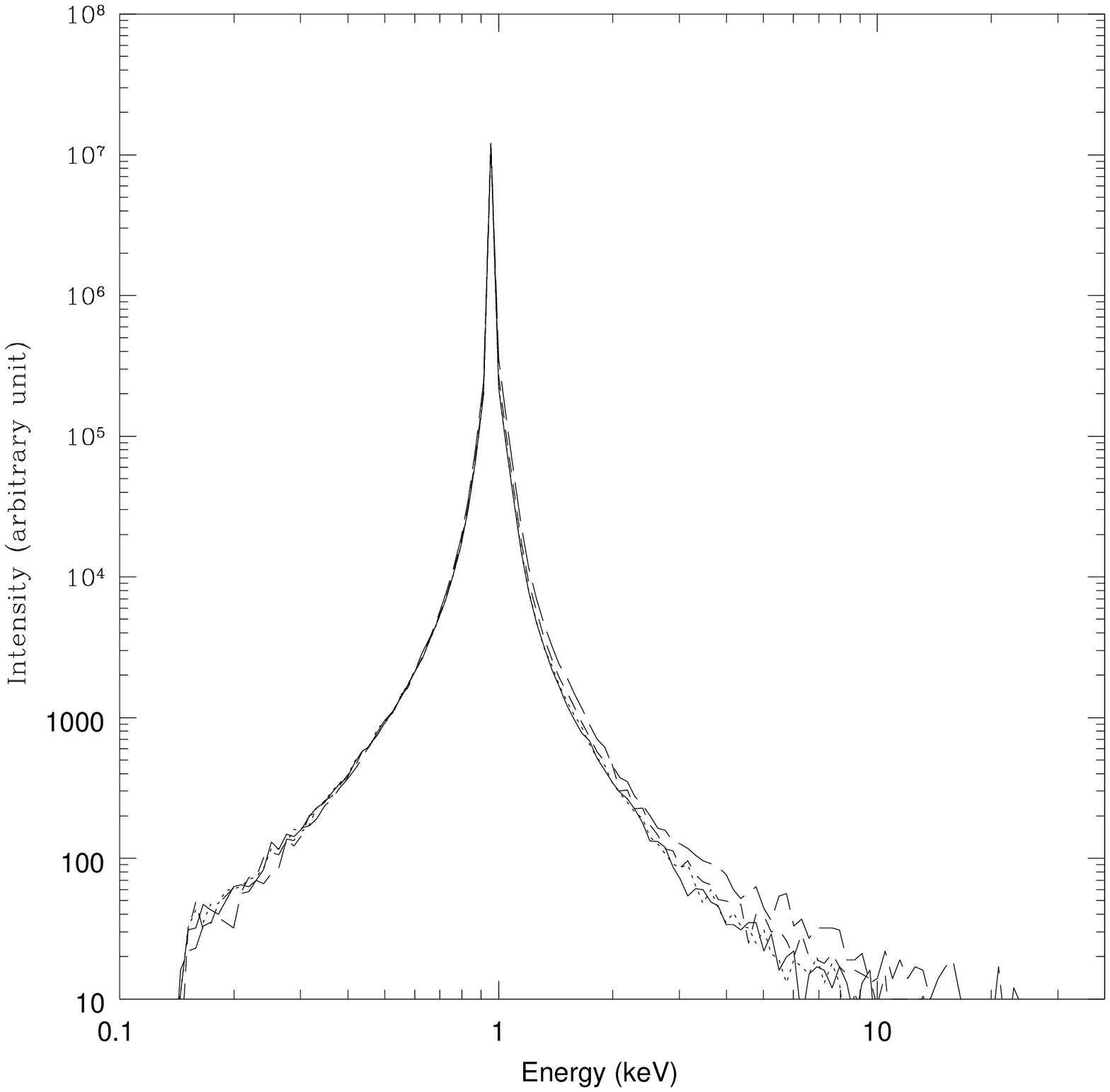} \qquad
            \includegraphics[width=5cm]{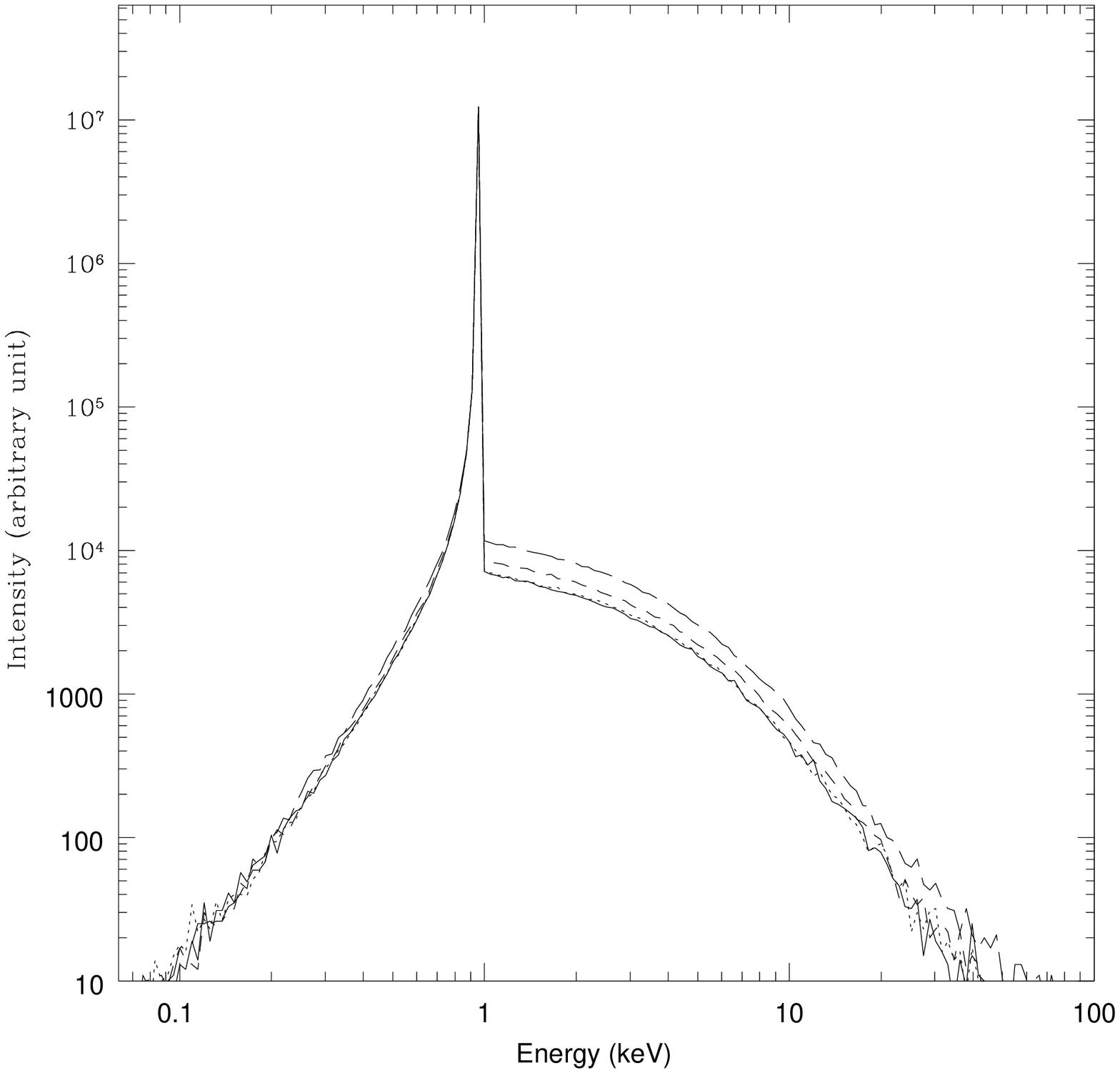}}

(a) \hskip 10.0cm (b)
\caption{Monte-Carlo Simulation Spectrum for \ee (a) and \ep (b). 
Curves are for spherical flow or $\phi=90^{\circ}$ (solid), $\phi=60^{\circ}$ (dotted), $\phi=30^{\circ}$ (dashed), $\phi=10^{\circ}$ (long dashed). $\phi$ is the angle shown in Fig. 5b. \label{f:f1}}
\end{figure}
As a representative case, let us compare the radiative properties of
fluids of two different composition \ie \ee and \ep, but starting with same energy at infinity
(${\cal E}_{\infty}=1.001$). The seed photons are mono-energetic photons of $1$KeV, and are randomly
injected through out the flow.
Various flow geometries from purely spherical (Fig. 5a) to conical
shaped (Fig. 5b) accretion has been considered. The spectrum for $\xi=0$ fluid is presented in Fig. 6a, while that of $\xi=1$ in Fig. 6b. The spectra from all the models (see Figs 5a-b) of accretion for \ee fluid is soft, and almost indistinguishable. However,
the spectra of the \ep fluid is hard and depends on the opening angle of the flow. This is expected since \ep fluid is hotter
than \ee fluid for the same outer boundary condition and hence electrons in \ep fluid more efficiently transfer energy to photon due to inverse Comptonization.
In presence of realistic cooling processes, the flow may become two temperature. We have also 
solved the two temperature solution for the \ep fluid in presence of bremsstrahlung and its Comptonization. The dependence of the flow solution on
the accretion rate is quite distinct. In Fig. 7, the proton ($T_p$) and electron ($T_e$) temperatures
are plotted for adiabatic flow, and flows with cooling process for accretion rates \ie
${\dot M}=0.1 ~\& 2.5$ ${\dot M}_{Edd}$.
The flows have the same outer boundary condition \ie ${\cal E}=1.000136$ at $r=10^4r_g$.
\begin{figure}[h]
\centerline{\includegraphics[width=8cm]{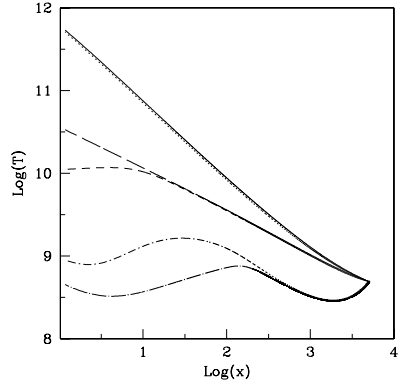} }
\vskip -1.0cm
\caption{Two temperature solution for \ep fluid. $T_p$ (solid) and $T_e$ (long dashed) for adiabatic flow,
$T_p$ (dotted) and $T_e$ (short dashed) when ${\dot M}=0.1 {\dot M}_{\rm Edd}$, and
$T_p$ (dot-short dashed) and $T_e$ (dot-long dashed) when ${\dot M}=2.5 {\dot M}_{\rm Edd}$.
The cooling process is Bremsstrahlung and Comptonization (inverse).}
\end{figure}

\section{Conclusion}
It has been shown in the paper that relativistic equation of state is very important to study relativistic
astrophysics of jets or accretion disc. It has also been shown that the composition of the fluid is extremely important too. Purely leptonic flow onto black holes do not show high energy phenomena like
shocks. Since transonic pair plasma fluid is not very relativistic, therefore spectra
from such flow are softer. If one considers fluid which are a mixture of leptons and hadrons,
then single temperature solution in absence of ohmic dissipation, may be an over simplified assumption. 
\section*{Acknowledgments}

DR was supported by the National Research Foundation of Korea through grant 2007-0093860.

%

%
%
\end{document}